\documentclass[pre,twocolumn,amsfonts,amssymb,amsmath,showpacs,floatfix]{revtex4}
\usepackage{enumerate}
\usepackage{amsmath}
\usepackage{amssymb}
\usepackage{graphicx}
\usepackage{psfrag}

\begin{document}

\title{Disordered asymmetric simple exclusion process: mean-field treatment}
\author{R.\ J.\ Harris}
\email{harris@thphys.ox.ac.uk}
\author{R.\ B.\ Stinchcombe}
\email{stinch@thphys.ox.ac.uk}
\affiliation{Theoretical Physics, Oxford University, 1 Keble Road, Oxford OX1 3NP, United Kingdom}
\date{\today}

\begin{abstract}

We provide two complementary approaches to the treatment of disorder in a fundamental nonequilibrium model, the asymmetric simple exclusion process.  Firstly, a mean-field steady state mapping is generalized to the disordered case, where it provides a mapping of probability distributions and demonstrates how disorder results in a new flat regime in the steady state current--density plot for periodic boundary conditions. This effect was earlier observed by Tripathy and Barma~\cite{Tripathy98} but we provide treatment for more general distributions of disorder, including both numerical results and analytic expressions for the width $2\Delta_C$ of the flat section.  We then apply an argument based on moving shock fronts~\cite{Popkov99} to show how this leads to an increase in the high current region of the phase diagram for open boundary conditions.   Secondly, we show how equivalent results can be obtained easily by taking the continuum limit of the problem and then using a disordered version of the well-known Cole--Hopf mapping to linearize the equation.  Within this approach we show that adding disorder induces a localization transformation (verified by numerical scaling), and $\Delta_C$ maps to an inverse localization length, helping to give a new physical interpretation to the problem.  
% In addition, this mapping to a linear problem allows us to discuss the effects of disorder on the full dynamics of the original problem.

\end{abstract}


\maketitle

\section{Introduction}

Many ``real-life'' nonequilibrium situations contain some kind of randomness/disorder and empirical observations for example in traffic flow illustrate that such disorder can lead to interesting new phenomena.  In nonequilibrium statistical mechanics even one-dimensional (1D) models can exhibit phase transitions (see e.g., the review by Evans~\cite{Evans00}) and we are particularly interested in the effects of disorder on these transitions.  Studies based on simple lattice based exclusion models incorporate collective effects while offering possibilities for analytic progress and easy computer simulation.

In this paper we concentrate on the effect of quenched substitutional disorder on one such lattice model---the well-known asymmetric simple exclusion model (ASEP).  The ASEP is one of the simplest nonequilibrium models with a boundary-driven steady state phase transition and thus plays a paradigmatic role in nonequilibrium statistical mechanics much as the Ising model does in the study of equilibrium systems.  The present work is entirely within the framework of a mean-field approximation and largely for the steady state but already shows many interesting effects on the phase transition such as an altered phase diagram and the presence of ``Griffiths phases''~\cite{Griffiths69}.  We hope subsequently to extend and compare this study with treatments allowing for fluctuation effects.  Previous approaches to disorder in the ASEP, and related models, can be found in the work of Krug~\cite{Krug00}, Kolwankar~\cite{Kolwankar00}, and others.   Furthermore, field-theoretic approaches which retain the fluctuations can be applied to higher-dimensional generalizations of the continuum version of the ASEP~\cite{Janssen86, Schmittmann95}.

The paper is organized as follows.  In Section~\ref{s:model} we define the model and summarize relevant results for the pure case.  In Section~\ref{s:approach} we outline our two main methods: a steady state mapping and a disordered generalization of the Cole--Hopf transformation.  These two approaches are then developed further in Sections~\ref{s:map} and~\ref{s:colehopf} respectively allowing us to characterize (both quantitatively and qualitatively) the effects of disorder.  In Section~\ref{s:fundpd} we show how this affects the current--density diagram for periodic boundary conditions and the phase diagram for open boundary conditions and discuss finite size effects.  Finally, in Section~\ref{s:conc}, we summarize our results and discuss areas for future work.

\section{The disordered asymmetric simple exclusion process}
\label{s:model}

\subsection{Definition of model}
\label{ss:model}

\begin{figure}
\begin{center}
\psfrag{A}[][]{{$\alpha$}}
\psfrag{B}[][]{{$\beta$}}
\psfrag{1}[Tc][Tc]{\scriptsize{1}}
\psfrag{2}[Tc][Tc]{\scriptsize{2}}
\psfrag{3}[Tc][Tc]{\scriptsize{3}}
\psfrag{l-1}[Tc][Tc]{\scriptsize{$l\!-\!1$}}
\psfrag{l}[Tc][Tc]{\scriptsize{$l$}}
\psfrag{l+1}[Tc][Tc]{\scriptsize{$l\!+\!1$}}
\psfrag{L}[Tc][Tc]{\scriptsize{$L$}}
\psfrag{pl}[Tc][Tc]{{$p_l$}}
\psfrag{ql}[Tc][Tc]{{$q_{l-1}$}}
\includegraphics*[width=0.8\columnwidth]{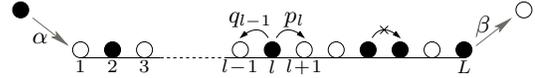}
\caption{DASEP with open boundary conditions.  Filled circles denote particles, open circles are vacancies.  The hard-core exclusion rule means the model incorporates collective effects.}
\label{f:DASEP}
\end{center}
\end{figure}

The general form of the bond-disordered asymmetric simple exclusion process (DASEP) is summarized by the schematic of Fig.~\ref{f:DASEP}.  
A particle at site $l$ hops to a \emph{vacant} nearest neighbor site on the right (left) with rate $p_l$ ($q_{l-1}$).  In a discrete time version of the model (as implemented in simulations), these rates are replaced by probabilities per time step and a random sequential update rule is applied.  Here we consider for simplicity only the totally asymmetric case, $q_l=0$; qualitatively similar results are expected in the partially asymmetric case.

Two obvious choices of boundary conditions are
\begin{itemize}

\item {
{\emph{Periodic boundary conditions}}: A particle from site $L$ can hop into a vacancy at site 1 with rate $p_L$.
}

\item{
{\emph{Open boundary conditions}}: Particles are injected at the left-hand end of the lattice and extracted at the right-hand end, forcing a particle current through the system.  
The usual convention is to insert particles onto site 1 with rate $\alpha$ if the site is empty and remove particles occupying site $L$ with rate $\beta$.  It is this case which is illustrated in Fig.~\ref{f:DASEP}.  

Another possibility (used for example in the work by Popkov and Sch{\"u}tz~\cite{Popkov99}) is to have fixed ``reservoir densities'' $\varrho^-$ and $\varrho^+$ respectively at left and right ends of the chain.  By current conservation at the boundaries we see that the correspondence between reservoir densities and input and output rates $\alpha$ and $\beta$ is given by
\begin{equation}
\alpha = p_0 \varrho^- \quad \beta=p_L (1-\varrho^+) 
\end{equation}
where $p_0$ is the hopping rate from the reservoir site 0 to the first proper site 1 and $p_L$ the rate from the last site $L$ to the reservoir at the right. 

In the pure case fixing reservoir densities is exactly equivalent to fixing input and output rates, but in the case where the $p_l$'s are disordered the two definitions are different.  We find both types of realization in representative problems e.g., traffic flow.

}

\end{itemize}

\subsection{Summary of results for pure case}

\begin{figure}
\begin{center}
\psfrag{J}{$J$}
\psfrag{r}{$\varrho$}
\psfrag{1}[][]{1}
\psfrag{0}[][]{0}
\psfrag{5}[Tc][Bc]{$\tfrac{1}{2}$}
\psfrag{p}[][]{$\tfrac{p}{4}$}
\includegraphics*[width=0.59\columnwidth]{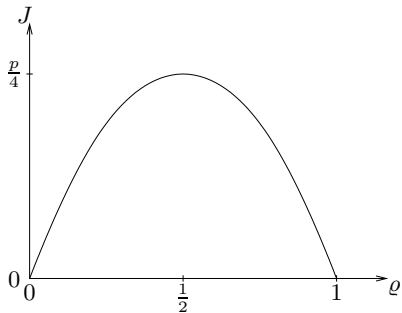}
\caption{Fundamental diagram (particle current $J$ versus density $\varrho$) for pure ASEP with periodic boundary conditions in the thermodynamic limit.  Note the particle--hole duality.}
\label{f:pfun}
\end{center}
\end{figure}

\begin{figure}
\begin{center}
\psfrag{J}[][]{High $J$}
\psfrag{p1}[][]{Low $\varrho$}
\psfrag{p2}[][]{High $\varrho$}
\psfrag{B}[Cr][Cr]{$\tfrac{\beta}{p}$}
\psfrag{A}[Tc][Bc]{$\tfrac{\alpha}{p}$}
\psfrag{5}[][]{$\tfrac{1}{2}$}
\psfrag{0}[][]{0}
\psfrag{1}[][]{1}
\includegraphics*[width=0.59\columnwidth]{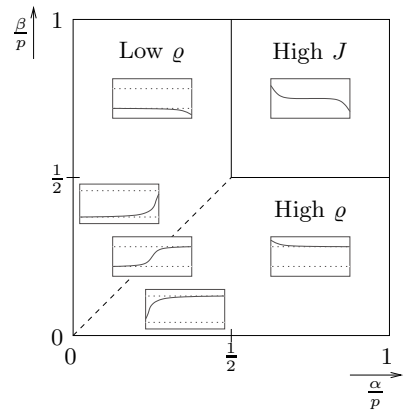}
\caption{Phase diagram for pure ASEP with open boundary conditions.  The second-order transition between high current ($\alpha/p > 1/2$, $\beta/p > 1/2$) and low current phases is represented by a solid line, while the first-order transition is shown dashed.   For an infinite system the current in the high current phase is $p/4$; for a finite size system a slightly larger current can be sustained.  The insets show sketches of the typical density profiles (site density $\varrho_l$ against lattice site $l$) in each region.}
\label{f:purepd}
\end{center}
\end{figure}

The pure ASEP with site-independent hopping rates (i.e., $p_l = p$ for all $l$) can be treated by a variety of approaches including recursive techniques~\cite{Schutz93c}, the steady state operator algebra formalism of Derrida \emph{et al.}~\cite{Derrida93b} and its dynamic generalization~\cite{Stinchcombe01}, or mapping to a quantum-spin system~\cite{Gwa92b}.  Recall that the exact solution for the steady state can be summarized by a current--density plot (the ``fundamental diagram'') for periodic boundary conditions and a phase diagram for open boundary conditions.  For later reference these are shown in Figs.~\ref{f:pfun} and~\ref{f:purepd} respectively.

A mean-field approximation reproduces the exact phase diagram and also gives the essence of the dynamics through a treatment based on moving shock fronts.  However, as might be expected from the low dimensionality, it gives incorrect values for the two static exponents and the dynamic exponent.  Similarly, we expect that a mean-field approach to the disordered problem will elucidate crucial features which are not controlled by fluctuations, such as modifications to the phase diagram and the dynamics but \emph{not} the critical behavior.  

Recent work by Enaud and Derrida~\cite{Enaud04} has looked at the affect of disorder on the first-order phase transition, in this paper we concentrate mainly on the affect of disorder on the second-order transition between low and high current phases.

\section{Two parallel approaches}
\label{s:approach}

In this section we outline our two basic approaches, explaining their use in the pure case and indicating how we extend them to treat disorder.

\subsection{Steady state mapping}
\label{ss:mapint}

For general quenched substitutional disorder we can use our knowledge of the hopping rules to write an exact expression for the average current across the bond between sites $l$ and $l+1$:
\begin{equation}
J_{l,l+1} = \langle p_l \, n_l (1 - n_{l+1})\rangle \label{e:Jdis}
\end{equation}
where $n_l=1$ or 0 for a particle of vacancy at site $l$ and the angular brackets denote an average over histories for a fixed $p_l$.
We introduce the average density for each site,
$
\langle n_l \rangle = \varrho_l,
$
and in the mean-field approximation ignore correlations between sites so
\begin{equation}
\langle n_l n_{l+1} \rangle \Rightarrow \langle n_l \rangle \langle n_{l+1} \rangle = \varrho_l \varrho_{l+1}.
\end{equation}
When the system has reached a steady state the densities are constant in time and hence from the continuity equation the current must be constant in space i.e., $J_{l,l+1}=J$ for all~$l$.  So, for the mean-field steady state we have
\begin{equation}
J = p_l \varrho_l (1 - \varrho_{l+1}) \label{e:J}
\end{equation}
which then gives a mapping for $\varrho_{l+1}$ in terms of $\varrho_l$:
\begin{equation}
\varrho_{l+1}=1-\frac{J}{p_l\varrho_l}. \label{e:map}
\end{equation}
For a particular realization of $p_l$'s, if we know $J$ and one of the densities (say $\varrho_1$) we can use this mapping to obtain the density profile for the whole system.  As we shall discuss in detail below, $J$ is limited by the requirement that for all $l$ we must have $0 \le \varrho_l \le 1$.

Note that we can also rearrange~\eqref{e:J} to give a mapping for $\varrho_l$ in terms of $\varrho_{l+1}$.  In terms of the hole density, $\sigma_l = 1 - \varrho_l$, this ``backward'' mapping is
\begin{equation}
\sigma_l = 1-\frac{J}{p_l\sigma_{l+1}}. \label{e:sigmap}
\end{equation}
which has exactly the same form as equation~\eqref{e:map} due to the particle--hole duality of the system.  

The mappings for the pure case where $p_l = p$ independent of $l$ have been given previously~\cite{Derrida92}.  There one finds that for low currents, $J<p/4$, the mapping has two fixed points.  Mapping in the direction of increasing $l$ the fixed point with higher $\varrho$ is stable and the lower one is unstable.  It is clear from~\eqref{e:sigmap} that for mapping in the opposite direction the stability of the fixed points is reversed and the high $\sigma$ (low $\varrho$) one is stable.  The resolution of this apparent paradox is simply that the steady state selection in a given case is determined by the boundary conditions.  Among the possible profiles are ``kink'' type solutions of the steady state pure mean-field profile map, having the form 
\begin{equation}
\varrho_l = \tfrac{1}{2} + \tfrac{1}{2} \tanh \phi \tanh(l \phi + \theta) \label{e:mfproflow}
\end{equation}
where $\tanh\phi = \sqrt{(1-4J/p)}$ and $\theta$ are determined by boundary conditions. 
 At $J=p/4$ the two fixed points combine in a half-stable fixed point at $\varrho=1/2$.  
In the high current regime, $J \ge p/4$, the mapping has no fixed points and density profiles have the form
\begin{equation}
\varrho_l = \tfrac{1}{2} - \tfrac{1}{2} \tan \phi' \tan(l \phi' +\theta') \label{e:mfprofhigh}
\end{equation}
with $\tan\phi' = \sqrt{(4J/p-1)}$.  Since we must have $0 \le \varrho_l \le 1$ for all $l$ then \eqref{e:mfprofhigh} clearly applies with non-zero $\phi'$ (i.e., $J > p/4$) only if $l$ is confined within the boundaries of a finite system.
These mean-field profiles agree qualitatively with exact solutions~\cite{Derrida93b} although the mean-field versions over exaggerate the sharpness of the shock front, which in practice is broadened by fluctuations.

\begin{figure}
\begin{center}
\psfrag{l}[][]{$l$}
\psfrag{r}[Bc][Tc]{$\varrho$}
\psfrag{0}[Tc][Tc]{\footnotesize{0}}
\psfrag{10}[Tc][Tc]{\footnotesize{10}}
\psfrag{20}[Tc][Tc]{\footnotesize{20}}
\psfrag{30}[Tc][Tc]{\footnotesize{30}}
\psfrag{40}[Tc][Tc]{\footnotesize{40}}
\psfrag{50}[Tc][Tc]{\footnotesize{50}}
\psfrag{60}[Tc][Tc]{\footnotesize{60}}
\psfrag{70}[Tc][Tc]{\footnotesize{70}}
\psfrag{80}[Tc][Tc]{\footnotesize{80}}
\psfrag{90}[Tc][Tc]{\footnotesize{90}}
\psfrag{100}[Tc][Tc]{\footnotesize{100}}
\psfrag{0.1}[Cr][Cr]{\footnotesize{0.1}}
\psfrag{0.2}[Cr][Cr]{\footnotesize{0.2}}
\psfrag{0.3}[Cr][Cr]{\footnotesize{0.3}}
\psfrag{0.4}[Cr][Cr]{\footnotesize{0.4}}
\psfrag{0.5}[Cr][Cr]{\footnotesize{0.5}}
\psfrag{0.6}[Cr][Cr]{\footnotesize{0.6}}
\psfrag{0.7}[Cr][Cr]{\footnotesize{0.7}}
\psfrag{0.8}[Cr][Cr]{\footnotesize{0.8}}
\psfrag{0.9}[Cr][Cr]{\footnotesize{0.9}}
\psfrag{mf mapping}[Cr][Cr]{\scriptsize{mf mapping}}
\psfrag{"mfprofmid.dat"}[Cr][Cr]{\scriptsize{mf mapping}}
\psfrag{MC data}[Cr][Cr]{\scriptsize{MC data}}
\psfrag{"profile350.dat"}[Cr][Cr]{\scriptsize{MC data}}
\includegraphics*[width=1.0\columnwidth]{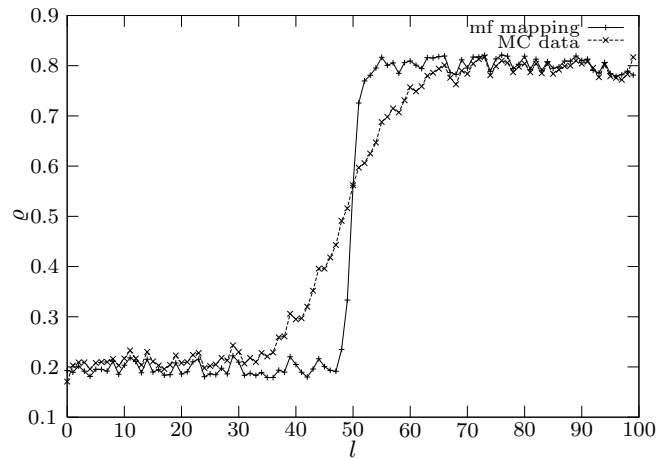}
\caption{Comparison of mean-field mapping and simulation data for densities in a system of 100 lattice points with open boundary conditions ($p_l$'s drawn from uniform distribution 0.45--0.55, simulation with $\alpha=\beta=0.2$). 
}
\label{f:mapsim}
\end{center}
\end{figure}

In the disordered case it is straightforward to iterate \eqref{e:map} by computer.  For specific realizations of disorder (i.e., particular choices of $\{ p_l \}$) we have compared densities from this mean-field mapping with profiles obtained by Monte Carlo simulation.  As shown in the low current example of Fig.~\ref{f:mapsim} 
there is a reasonable qualitative fit though again the mean-field shock front is sharper than the simulation result.  Close to the pure critical current one sees that in some regions (corresponding to groups of ``weak'' bonds with low $p_l$) the profile has roughly the high current form while in other regions (``strong'' bonds) it takes the low current form---this is the analogue of Griffith's phases in a magnet~\cite{Griffiths69}.  We performed comparisons for the open boundary case (see~\cite{Tripathy98} for similar discussion with periodic boundary conditions) where the simulation is controlled by parameters $\alpha$ and $\beta$.  A computer fitting procedure was used to match $\alpha$ and $\beta$ to the value of $J$ and one of the $\varrho_l$s needed to implement the mean-field mapping (for the pure case this can be done exactly). 

Such examples for specific realizations of disorder support the qualitative validity of the mean-field approach.  In Section~\ref{s:map}  we develop a more general analysis to predict the typical effects of a given distribution of disorder (Gaussian, uniform etc.).  By considering the mapping of density distributions we are able to show that a characteristic effect of disorder is a shift in the average density.  The physical meaning is explored further in Section~\ref{s:fundpd}.

\subsection{Continuum limit and Cole--Hopf transformation}
\label{ss:chintro}

Here we consider the continuum limit of the DASEP and introduce a disordered generalization of the well-known Cole--Hopf transformation~\cite{Hopf50,Cole51}.  

For ``smooth enough'' disorder (and working once again in the mean-field approximation) we can take the continuum limit of equation~\eqref{e:Jdis}, to arrive at
\begin{equation}
J=p(x)\left(\varrho(1-\varrho)-\frac{1}{2}\frac{\partial\varrho}{\partial x}\right) \label{e:Jcont}
\end{equation}
where $J$ and $\varrho$ are in general functions of continuous position $x$ and time $t$, and we set the lattice spacing equal to 1 for convenience.  
Substituting this into the continuity equation yields a (noiseless) disordered Burger's type equation
\begin{equation}
\frac{\partial \varrho}{\partial t} = -\frac{\partial}{\partial x} \left[ p(x)\left(\varrho(1-\varrho)-\frac{1}{2}\frac{\partial\varrho}{\partial x}\right) \right].
\end{equation}
The next step is to transform to a height variable $h$, such that $\partial h / \partial x = \varrho - 1/2$, giving 
\begin{equation}
\frac{\partial^2 h}{\partial x \partial t} = -\frac{\partial}{\partial x} \left[ p(x)\left( \frac{1}{4} - \left(\frac{\partial h}{\partial x} \right)^2 - \frac{1}{2}\frac{\partial^2 h}{\partial x^2} \right) \right].
\end{equation}
This can be trivially integrated with respect to $x$ to give a noiseless disordered version of the growth model studied by Kardar, Parisi and Zhang~\cite{Kardar86}.  We then put $h(x,t) = \lambda \ln[u(x,t)] + f(t)$ and choose the arbitrary function $f(t)$ and the constant $\lambda$ so as to remove all non-linear terms.  This disordered generalization of the Cole--Hopf transformation finally gives us
\begin{equation}
\frac{\partial u}{\partial t} = D(x) \frac{\partial^2 u}{\partial x^2} - D(x) u \label{e:disdiff}
\end{equation}
where $D(x)=p(x)/2$.  Equation~\eqref{e:disdiff} is a linear equation (therefore much easier to treat numerically and analytically) which still preserves the full dynamics of the system.  From its solution $\varrho$ is given by the inverse Cole--Hopf transformation
\begin{equation}
\varrho - \frac{1}{2} = \frac{1}{2} \frac{ \partial \ln u}{\partial x}. \label{e:invCH}
\end{equation} 
For a steady state solution for $\varrho$ then $u$ must be separable with time dependence $e^{-\omega t}$ and $\omega=2J$.

In the pure case the second term on the right-hand side of~\eqref{e:disdiff} can be absorbed into the definition of the transformation to leave us with just the diffusion equation,
\begin{equation}
\frac{\partial u}{\partial t} = D \frac{\partial^2 u}{\partial x^2}
\end{equation}
This diffusion equation can be trivially solved and mapping back through the Cole--Hopf transformation then gives the well known pure continuum steady state solutions which are the continuum versions of~\eqref{e:mfproflow} and~\eqref{e:mfprofhigh}.

For the disordered case, the situation is more complicated and in particular we have to include the $-D(x) u$ term on the right-hand side of~\eqref{e:disdiff}.  It is not immediately obvious how to treat this equation for general $D(x)$, although it might be possible to solve it for specific $D(x)$ or to do some kind of WKB type approximation.  In Section~\ref{s:colehopf} we shall show how much useful information can be obtained via a more powerful numerical scaling approach which reveals that disorder induces a localization transition.  This provides a complementary approach to the steady state mapping outlined above and enables us to interpret the effect of disorder as a localization transition in the transformed system. 

\section{Steady state mean-field mapping for disordered case}
\label{s:map}

\subsection{Mapping of distributions}
\label{ss:mapdis}

Here we return to look in detail at the steady state mapping with a known distribution of disorder.  It is convenient to take the disordered variable as
\begin{equation}
\gamma_l= J/p_{l}.
\end{equation}
If the position-independent distribution $f(\gamma_l$) is known, then one can use the mapping~\eqref{e:map} to relate the probability distribution $w$ of $\varrho_{l+1}$ to the distribution of $\varrho_l$:
\begin{equation}
w_{l+1}(\varrho_{l+1}) = \int w_l\!\left(\frac{\gamma}{1-\varrho_{l+1}}\right)\! \frac{\gamma_l}{(1-\varrho_{l+1})^2} f(\gamma) \, d\gamma. \label{e:mapdistorig}
\end{equation}
The subscripts on the $w$'s in~\eqref{e:mapdistorig} indicate that we expect the distribution to change as we map through the system.  For example, if we start from a known $\varrho_1$ (i.e., $w_1$ is a delta function) then the width of the distribution will obviously increase as we look at $w_2$, $w_3$ etc.

After iterating the mapping for many steps the density distribution will eventually converge on some fixed point shape, the position of which will depend on the direction in which we map (just as in the pure case discussion in~\ref{ss:mapint}).    This stationary probability density is the distribution of $\varrho_l$'s which we would expect to see in the periodic boundary case.   However, as we shall demonstrate in~\ref{ss:pd}, by considering moving shock type solutions we can also gain some information about the expected open boundary profiles and phase diagram.

Numerically therefore, we look for the stationary probability density of $\varrho_l$'s for different $J$ (averaged over many realizations of disorder).  In practice this involves using the mapping of equation~\eqref{e:map} for large system sizes (say 10,000 lattice points), repeating for different $\varrho_0$ and different realizations of disorder then creating a histogram of the values of $\varrho$.  

\begin{figure}
\begin{center}
\psfrag{J}[][]{$J$}
\psfrag{D}[Bc][Tc]{$\langle \varrho - 1/2 \rangle$}
\psfrag{0.08}[Tc][Tc]{\footnotesize{0.08}}
\psfrag{0.09}[Tc][Tc]{\footnotesize{0.09}}
\psfrag{0.1b}[Tc][Tc]{\footnotesize{0.10}}
\psfrag{0.11}[Tc][Tc]{\footnotesize{0.11}}
\psfrag{0.12}[Tc][Tc]{\footnotesize{0.12}}
\psfrag{0.13}[Tc][Tc]{\footnotesize{0.13}}
\psfrag{0.14}[Tc][Tc]{\footnotesize{0.14}}
\psfrag{0.15}[Tc][Tc]{\footnotesize{0.15}}
\psfrag{-0.05}[Cr][Cr]{\footnotesize{-0.05}}
\psfrag{0}[Cr][Cr]{\footnotesize{0.00}}
\psfrag{0.05}[Cr][Cr]{\footnotesize{0.05}}
\psfrag{0.1}[Cr][Cr]{\footnotesize{0.10}}
\psfrag{0.15}[Cr][Cr]{\footnotesize{0.15}}
\psfrag{0.2}[Cr][Cr]{\footnotesize{0.20}}
\psfrag{0.25}[Cr][Cr]{\footnotesize{0.25}}
\psfrag{0.3}[Cr][Cr]{\footnotesize{0.30}}
\psfrag{"mapeprangeJrho2.dat"}[Cr][Cr]{\scriptsize{pure, $\sigma_\eta=0$}}
\psfrag{"mapegstrangeJrho2.dat"}[Cr][Cr]{\scriptsize{disordered, $\sigma_\eta=0.2$}} %replace with longer run?
\psfrag{"mapegDDrangeJrho2.dat"}[Cr][Cr]{\scriptsize{disordered, $\sigma_\eta=0.5$}}
\psfrag{"mapegwkrangeJrho2.dat"}[Cr][Cr]{\scriptsize{disordered, $\sigma_\eta=0.02$}}
\includegraphics*[width=1.0\columnwidth]{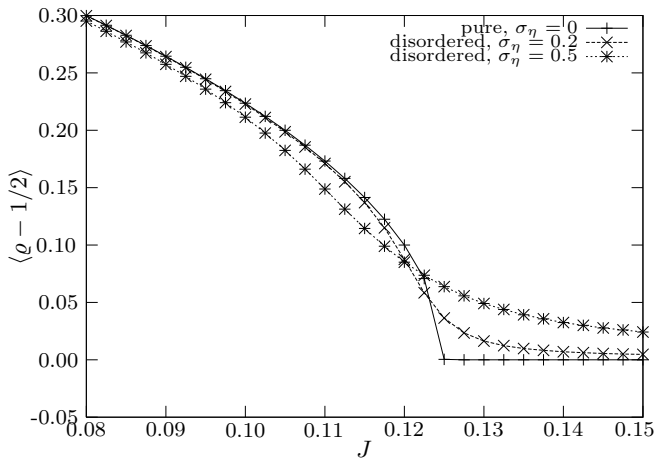}
\caption{Numerical mapping data for $\Delta \equiv \langle \varrho-1/2 \rangle$ versus $J$ in pure case and specimen disordered cases.  Pure case has $\bar{\eta}=2.0$ ($\eta$ defined as $1/p$) corresponding to a pure critical current $J_C^0$ of 0.125, disordered cases are Gaussian distributions with the same mean and standard deviations $\sigma_\eta=0.2,0.5$.  Lines are provided as an aid to the eye.  For arbitrarily weak disorder, $\Delta$ is non-zero for all $J$.}
\label{f:distnum}
\end{center}
\end{figure}

We find that a convenient order parameter to characterize the distributions is the asymmetry about $\varrho=1/2$ given by $\Delta \equiv \langle \varrho - 1/2 \rangle$ (where numerically we take the average over all $\varrho$ in the physically accessible region between 0 and 1).  In the pure case we find that $\Delta$ is zero in the high current regime and non-zero in the low current regime (see Fig.~\ref{f:distnum}).  This is just what we would expect from using the pure mapping (see again~\ref{ss:mapint}) in the forward direction: in the low current phase the densities in the bulk of the system will approach the upper fixed point, whereas for a long system in the high current phase the densities in the bulk are all very close to $1/2$.  However, in the disordered case we find that  $\Delta$ is non-zero for all $J$ as shown for typical examples in Fig.~\ref{f:distnum}. 

We now consider an analytical argument to reproduce this data basing our method on the work of Hirota~\cite{Hirota73} for different random mapping processes.
In our case, we impose the fixed point condition by setting $w_{l}=w_{l+1}=w$ so that~\eqref{e:mapdistorig} gives (dropping redundant subscripts)
\begin{equation}
w(\varrho) = \int w\!\left(\frac{\gamma}{1-\varrho}\right)\! \frac{\gamma}{(1-\varrho)^2} f(\gamma) \, d\gamma. \label{e:int}
\end{equation}

This integral equation is difficult to solve analytically but we can find approximate solutions by considering the dominant terms in different regimes.  In~\ref{ss:above}--\ref{ss:crit} we pursue this approach above, below and close to the pure critical point.  Then in~\ref{ss:lorentz} we consider the specific soluble example of a Lorentzian distribution of disorder.

\subsection{Calculation of $\Delta$ above pure critical point}
\label{ss:above}

Above the pure critical point the integral in~\eqref{e:int} is dominated by the peak in $f(\gamma)$.  For disorder sharply peaked about $\bar{\gamma}$ with small variance $\sigma_\gamma^2$, we can perform an expansion of the integrand in powers of $\sigma_{\gamma}^2$ and obtain a functional differential equation
\begin{equation}
w(\varrho) = w\!\left(\frac{\bar{\gamma}}{1-\varrho} \right)\! \frac{\bar{\gamma}}{(1-\varrho)^2} + \frac{\sigma_\gamma^2}{2} \frac{d^2}{d\varrho^2} \left[ w \left( \frac{\bar{\gamma}}{1-\varrho} \right) \right] \label{e:func}
\end{equation}
where terms involving higher moments have been neglected.  We now wish to solve~\eqref{e:func} for normalized non-negative $w(\varrho)$.  Note that in order to be able to carry out the integrals analytically we here allow $\varrho$ to take any real value whereas in the physical problem $0 \le \varrho \le 1$.  As we shall discuss later this is not expected to introduce much of an error providing $J$ is not too high compared with the pure critical current.  For $\sigma_{\gamma}^2$ ``small'' (in a sense to be clarified) we assume a solution of the form
\begin{equation}
w(\varrho) = w_0(\varrho) + \frac{\sigma_\gamma^2}{2} w_1 (\varrho). \label{e:perturb}
\end{equation}
The pure solution $w_0(\varrho)$ is easily shown to be
\begin{equation}
w_0(\varrho) = A \frac{1}{\varrho^2 - \varrho + \bar{\gamma}} \label{e:w0}
\end{equation}
with $A$ a normalization constant given by $\sqrt{\bar{\gamma}-1/4}/\pi$.  This solution is only valid for $\bar{\gamma} > 1/4$ which corresponds to being in the high current phase of the corresponding pure model.   Note that we can also get this result from the known high current mean-field pure result~\eqref{e:mfprofhigh} using the obvious relationship $w(\varrho) \sim 1/|\frac{d\varrho}{dl}|$.  By considering this form for $w_0(\varrho)$, we see that to satisfy~\eqref{e:func} $w_1(\varrho)$ must have a factor $(\varrho^2-\varrho+\bar{\gamma})^3$ in the denominator.  Straightforward calculation gives
\begin{equation}
w_1(\varrho) = \frac{2A}{\bar{\gamma}} \frac{\varrho^3 - \bar{\gamma} \varrho}{(\varrho^2 - \varrho + \bar{\gamma})^3}.  \label{e:w1rho}
\end{equation}
From this expression for the stationary probability distribution we can calculate (via contour integration) the average value of $\varrho - 1/2$:
\begin{align}
\Delta % & \equiv \left\langle \varrho-\tfrac{1}{2} \right\rangle \\
&= \int_{-\infty}^\infty \left( \varrho - \tfrac{1}{2} \right) w(\varrho) \, d\varrho  \label{e:hcresint} \\
&= \frac{\sigma_\gamma^2}{4(\bar{\gamma} - \tfrac{1}{4})}. \label{e:hcres}
\end{align}
This is the chief analytical result of this subsection; we now compare it with numerics and discuss its validity.

\begin{figure}
\begin{center}
\psfrag{J}[][]{$J$}
\psfrag{D}[Bc][Tc]{$\langle \varrho - 1/2 \rangle$}
\psfrag{0.125}[Tc][Tc]{\footnotesize{0.125}}
\psfrag{0.13}[Tc][Tc]{\footnotesize{0.130}}
\psfrag{0.135}[Tc][Tc]{\footnotesize{0.135}}
\psfrag{0.14}[Tc][Tc]{\footnotesize{0.140}}
\psfrag{0.145}[Tc][Tc]{\footnotesize{0.145}}
\psfrag{0.15}[Tc][Tc]{\footnotesize{0.150}}
\psfrag{0}[Cr][Cr]{\footnotesize{0}}
\psfrag{0.005}[Cr][Cr]{\footnotesize{0.005}}
\psfrag{0.01}[Cr][Cr]{\footnotesize{0.010}}
\psfrag{0.015}[Cr][Cr]{\footnotesize{0.015}}
\psfrag{0.02}[Cr][Cr]{\footnotesize{0.020}}
\psfrag{0.025}[Cr][Cr]{\footnotesize{0.025}}
\psfrag{0.03}[Cr][Cr]{\footnotesize{0.030}}
\psfrag{0.035}[Cr][Cr]{\footnotesize{0.035}}
\psfrag{0.04}[Cr][Cr]{\footnotesize{0.040}}
\psfrag{0.045}[Cr][Cr]{\footnotesize{0.045}}
\psfrag{0.05}[Cr][Cr]{\footnotesize{0.050}}
\psfrag{f(x)}[Cr][Cr]{\scriptsize{analytical prediction of~\eqref{e:hcresJ}}}
\psfrag{"mapegstrangeJrho2.dat"}[Cr][Cr]{\scriptsize{numerical mapping data}}
\includegraphics*[width=1.0\columnwidth]{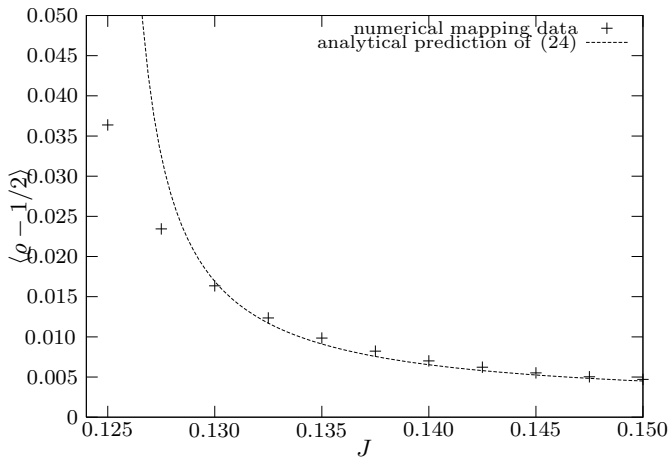}
\caption{Comparison of numerical and analytical results for $\Delta$, above pure critical current ($J_C^0=0.125$).  Numerical data for Gaussian disorder with $\bar{\eta}=2.0$ and $\sigma_\eta=0.2$.  For weaker disorder the analytic expression is valid for currents closer to the pure critical point---see discussion in text.} 
\label{f:hcres}
\end{center}
\end{figure}

For comparison with data it is more helpful to write~\eqref{e:hcres} in terms of the current $J$ and the mean $\bar{\eta}$ and variance $\sigma_\eta^2$ of the inverse hopping probability $\eta_l = 1/p_l$,
\begin{equation}
\Delta = \frac{J^2 \sigma_\eta^2}{4(J\bar{\eta} - \tfrac{1}{4})}. \label{e:hcresJ}
\end{equation}
In Fig.~\ref{f:hcres} we compare the prediction of this analytical result with the data from our numerical mapping with a Gaussian distribution of disorder. 
We see that~\eqref{e:hcresJ} reproduces well the general trend of the data but there are a couple of obvious problems.  For high values of $J$, the analytical expression slightly under-estimates the numerical result---this is due to the fact that analytically $-\infty<\varrho<\infty$ whereas in the numerics (as in the physical problem) we have averaged over $0 \leq \varrho \leq 1$.  As expected, this discrepancy becomes larger for high $J$ since then the density profile is steeper and is concentrated less about $\varrho \sim 1/2$ so the tails of the distribution are more important.  One can numerically integrate (e.g., using {\scshape{maple}} or similar) the expression in~\eqref{e:hcresint} between 0 and 1 (and adjust the normalization correspondingly) and obtain values in better agreement with the data.  

A more serious problem is the fact that our analytical expression for $\Delta$ tends to infinity as $J$ tends to the pure critical current $J^0_C = 1/(4\bar{\eta})$, whereas numerically we see no divergence in $\Delta$.  In fact it is easy to see why the analytical method fails close to the critical point.  As $J \rightarrow J_C^0$, the pure distribution $w_0(\varrho)$ becomes more and more sharply peaked about $\varrho = 1/2$, however $w_1(\varrho)$ will be even more sharply peaked (due to the cubed term in the denominator) and eventually the magnitude of this first order correction becomes large compared with $w_0(\varrho)$ for some values of $\varrho$.  This leads to unphysical results such as negative probability densities.  A rough calculation shows that in order for the perturbation expansion of~\eqref{e:func} and~\eqref{e:perturb} to be valid we require $\sigma_{\gamma}^2 \ll (\bar{\gamma} - 1/4)$ and this condition becomes impossible to satisfy as $\bar{\gamma} \rightarrow 1/4$.   Alternative approaches to obtain an expression valid near the critical point will be considered in~\ref{ss:crit} below.  

\subsection{$\Delta$ below pure critical point}

\begin{figure}
\begin{center}
\psfrag{J}[][]{$J$}
\psfrag{D}[Bc][Tc]{$\langle \varrho - 1/2 \rangle$}
\psfrag{0}[Tr][Tr]{\footnotesize{0}}
\psfrag{0.02}[Tc][Tc]{\footnotesize{0.02}}
\psfrag{0.04}[Tc][Tc]{\footnotesize{0.04}}
\psfrag{0.06}[Tc][Tc]{\footnotesize{0.06}}
\psfrag{0.08}[Tc][Tc]{\footnotesize{0.08}}
\psfrag{0.1b}[Tc][Tc]{\footnotesize{0.10}}
\psfrag{0.12}[Tc][Tc]{\footnotesize{0.12}}
\psfrag{0.05}[Cr][Cr]{\footnotesize{0.05}}
\psfrag{0.1}[Cr][Cr]{\footnotesize{0.10}}
\psfrag{0.15}[Cr][Cr]{\footnotesize{0.15}}
\psfrag{0.2}[Cr][Cr]{\footnotesize{0.20}}
\psfrag{0.25}[Cr][Cr]{\footnotesize{0.25}}
\psfrag{0.3}[Cr][Cr]{\footnotesize{0.30}}
\psfrag{0.35}[Cr][Cr]{\footnotesize{0.35}}
\psfrag{0.4}[Cr][Cr]{\footnotesize{0.40}}
\psfrag{0.45}[Cr][Cr]{\footnotesize{0.45}}
\psfrag{0.5}[Cr][Cr]{\footnotesize{0.50}}
\psfrag{g(x)}[Cr][Cr]{\scriptsize{analytical prediction of~\eqref{e:lcres}}}
\psfrag{"mapegstrangeJrho2.dat"}[Cr][Cr]{\scriptsize{numerical mapping data}}
\includegraphics*[width=1.0\columnwidth]{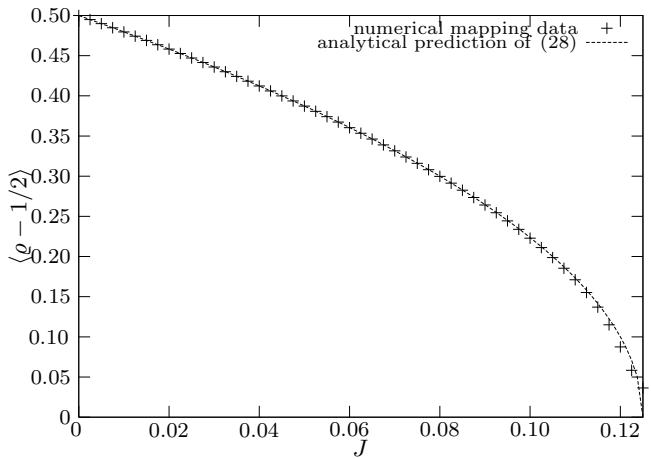}
\caption{Comparison of numerical and analytical results for $\Delta$, below pure critical current ($J_C^0=0.125$).  Numerical data for Gaussian disorder with $\bar{\eta}=2.0$ and $\sigma_\eta=0.2$.  Note the discrepancy close to $J_C^0$.}
\label{f:lcres}
\end{center}
\end{figure}

Far below the critical point the integral in~\eqref{e:int} is dominated by the contribution from the sharp peak of $w(\frac{\gamma}{1-\varrho})$.  Performing a saddle-point expansion about this peak (and for convenience assuming a Gaussian distribution for the disorder) we obtain
\begin{equation}
w(\varrho)=\frac{\tilde{\varrho} \left[ w(\tilde{\varrho}) \right] ^2 \, \exp \left( -\frac{\left[(1-\varrho)\tilde{\varrho} - \bar{\gamma}\right]^2}{2\sigma_\gamma^2}\right)}{\sigma \left( w''(\tilde{\varrho})w(\tilde{\varrho}) - \left[w'(\tilde{\varrho})\right]^2\right)^{1/2}}. \label{e:loww}
\end{equation}
Now in the saddle-point expansion $\tilde{\varrho}$ is defined as the maximum of $w(\varrho)$ so for consistency it must correspond to the maximum of the Gaussian in~\eqref{e:loww}, i.e., we have
\begin{equation}
(1-\tilde{\varrho})\tilde{\varrho}=\bar{\gamma}
\end{equation}
which is just the condition for the fixed-points of the pure low current mapping.  We recall that the upper of these two fixed points is stable as we map forward through the system while the lower one is unstable.  So as $\sigma_\gamma \rightarrow 0$, $w(\varrho)$ of equation~\eqref{e:loww} tends to a delta function about the upper fixed point.  The addition of disorder, broadens this pure distribution to a Gaussian with the same mean $\tilde{\varrho}$ and standard deviation $\sigma_\gamma/\tilde{\varrho}$, i.e.,
\begin{equation}
w(\varrho) = (2\pi)^{-1/2}(\tilde{\varrho}/\sigma_\gamma) \, \exp \left(-\frac{(\varrho-\tilde{\varrho})^2}{2(\sigma_\gamma/\varrho)^2} \right).
\end{equation}
Calculating $\Delta$ in this approximation is trivial and gives
\begin{equation}
\Delta = \sqrt{\tfrac{1}{4}-\bar{\gamma}}. \label{e:lcres}
\end{equation}
So we conclude that, for currents well below the pure critical current, disorder doesn't change the pure result for $\Delta$.  This agrees with the numerical results for $J \ll J^0_C$ shown in Fig.~\ref{f:lcres}.

Again our analytic prediction fails close to the critical point; this is due to the breakdown of the assumption that the major contribution to the integral equation~\eqref{e:int} is due to the sharp peak in $w(\varrho)$.  From the above analysis we can see that the standard deviation in $\gamma$ of the peak in $w(\frac{\gamma}{1-\varrho})$ is about $\sigma_\gamma (1-\tilde{\varrho})/\tilde{\varrho}$ so if this peak is to be sharper than the one in $f(\gamma)$ we require:
\begin{align}
\sigma_\gamma(1-\tilde{\varrho}) / \tilde{\varrho} &\ll \sigma_\gamma \\
\tilde{\varrho} &\gg \tfrac{1}{2}.
\end{align}
This condition breaks down as we approach the pure critical point and a more sophisticated analysis becomes necessary (see Subsection~\ref{ss:crit})

Note that in order to obtain an analytic approximation for $\Delta$ in the low current regime we needed to assume a particular distribution for the disorder.  Repeating the procedure for different distributions (e.g., uniform, binary) we find that $w(\varrho)$ has a different form in each case but to a first approximation $\Delta$ takes the pure value in all cases.  This is in contrast to the high current regime treated in~\ref{ss:above} where the exact form of disorder is irrelevant to first order (the relevant parameter is the standard deviation of $1/p$) and $w(\varrho)$ takes the universal form given by~\eqref{e:perturb}--\eqref{e:w1rho}.  The numerics confirms these arguments. 

\subsection{Calculation of $\Delta$ around pure critical point}
\label{ss:crit}

Ideally we would like an expression for $\Delta$ at and very close to the pure critical current since this is the most interesting regime physically (a large system cannot sustain currents much above the critical point as will be discussed in detail in Section~\ref{s:fundpd} below).  Unfortunately in this intermediate regime it is not easy to see how to treat~\eqref{e:int} as both $w(\frac{\gamma}{1-\varrho})$ and $f(\gamma)$ are sharply peaked.

One approach is to assume that the product $w(\frac{\gamma}{1-\varrho})f(\gamma)$ is sharply peaked in $\gamma$ and perform a saddle-point expansion on this.  For convenience, we define $y(\varrho) \equiv -\ln\left[ w(\varrho) \right]$ and consider a Gaussian distribution of disorder.  Then the saddle-point $\tilde{x}$  is defined by
\begin{equation}
0 = \frac{\left[ \tilde{x}(1-\varrho) - \bar{\gamma} \right]}{\sigma_\gamma^2}(1-\varrho) - \frac{1}{\tilde{x}} + y'(\tilde{x})
\end{equation}
where the prime denotes differentiation with respect to the argument of the function.
Evaluating~\eqref{e:int} about this saddle-point yields
\begin{equation}
y(\varrho) = \frac{\left[ \tilde{x}(1-\varrho) - \bar{\gamma} \right] ^2}{2 \sigma_\gamma^2} - \ln\tilde{x} + y(\tilde{x}).
\end{equation}
In principle these equations are sufficient to determine $\tilde{x}$ and $y(\varrho)$ but the implicit definition of $\tilde{x}$ makes further analytic progress extremely difficult.  Instead we consider, in the next section, a specific (Lorentzian) distribution of disorder where the mapping equations turn out to be exactly soluble.   This particular case helps build up a general picture of what happens near the critical point.  

\subsection{An exactly soluble case}
\label{ss:lorentz}

Hirota and Ishii~\cite{Hirota71} have treated \emph{exactly} the case where the disorder variable in their mapping has a Lorentzian distribution.  They show that the stationary probability density is also Lorentzian and calculate its width and mean.  Our mapping equation has a different form but is also amenable to analytical treatment for the case of Lorentzian disorder.  This is therefore a useful test case where we can compare numerics and analytics \emph{for the full range of $J$}.

If $\gamma$ is drawn from the Lorentzian distribution 
\begin{equation}
f(\gamma)=\frac{1}{\pi}\frac{\Gamma}{(\gamma-\bar{\gamma})^2+ \Gamma^2},
\end{equation}
and we assume a Lorentzian distribution for $\varrho_l$ (mean $\bar{\varrho}$, width $t$) 
then we can integrate exactly the integral mapping equation~\eqref{e:int} to find that $\varrho_{l+1}$ also obeys a Lorentzian distribution
with mean and width given by:
\begin{align}
\bar{\varrho}' &= 1 - \frac{\bar{\gamma}\bar{\varrho} - \Gamma t}{\bar{\varrho}^2 + t^2} \\
t' &= \frac{\Gamma\bar{\varrho} + \bar{\gamma} t}{\bar{\varrho}^2 + t^2}.
\end{align}

Following~\cite{Hirota71} we can characterize each Lorentzian distribution by a complex number whose real part represents the mean and whose imaginary part represents the the width.  Then the mapping relationship is
\begin{equation}
R' = 1 - G^\dagger / R
\end{equation}
where $R=\bar{\varrho}+it$, $G=\bar{\gamma}+i\Gamma$ and the dagger denotes complex conjugation.  
Now the fixed point of the mapping is given by the Lorentzian distribution characterized by $R^*=\bar{\varrho}^*+it^*$, with
\begin{align}
\bar{\varrho}^* &= \frac{t^* + \Gamma}{2t^*} \\
t^* &= x^{1/2}
\end{align}
where $x$ is the positive root of
\begin{equation}
x^2 + \left( \tfrac{1}{4} - \bar{\gamma} \right) x - \tfrac{1}{4} \Gamma^2 = 0.
\end{equation}

\begin{figure}
\begin{center}
\psfrag{J}[][]{$J$}
\psfrag{D}[Bc][Tc]{$\langle \varrho - 1/2 \rangle$}
\psfrag{0b}[Tc][Tc]{\footnotesize{0}}
\psfrag{0.05b}[Tc][Tc]{\footnotesize{0.05}}
\psfrag{0.1b}[Tc][Tc]{\footnotesize{0.10}}
\psfrag{0.15b}[Tc][Tc]{\footnotesize{0.15}}
\psfrag{0.2b}[Tc][Tc]{\footnotesize{0.20}}
\psfrag{0.25b}[Tc][Tc]{\footnotesize{0.25}}
\psfrag{0}[Cr][Cr]{\footnotesize{0}}
\psfrag{0.05}[Cr][Cr]{\footnotesize{0.05}}
\psfrag{0.1}[Cr][Cr]{\footnotesize{0.10}}
\psfrag{0.15}[Cr][Cr]{\footnotesize{0.15}}
\psfrag{0.2}[Cr][Cr]{\footnotesize{0.20}}
\psfrag{0.25}[Cr][Cr]{\footnotesize{0.25}}
\psfrag{0.3}[Cr][Cr]{\footnotesize{0.30}}
\psfrag{0.35}[Cr][Cr]{\footnotesize{0.35}}
\psfrag{0.4}[Cr][Cr]{\footnotesize{0.40}}
\psfrag{0.45}[Cr][Cr]{\footnotesize{0.45}}
\psfrag{0.5}[Cr][Cr]{\footnotesize{0.50}}
\psfrag{-0.1}[Tc][Tc]{\footnotesize{-0.10}}
\psfrag{g(mean*x, sd*x)}[Cr][Cr]{\scriptsize{analytical prediction of~\eqref{e:lorentzdel}}}
\psfrag{g2(mean*x, sd*x)+h(mean*x, sd*x)}[Cr][Cr]{\scriptsize{improved analytical prediction of~\eqref{e:lorentzdelimp}}}
\psfrag{"mapegstrspreadcJrho2.dat"}[Cr][Cr]{\scriptsize{numerical mapping data}}
\includegraphics*[width=1.0\columnwidth]{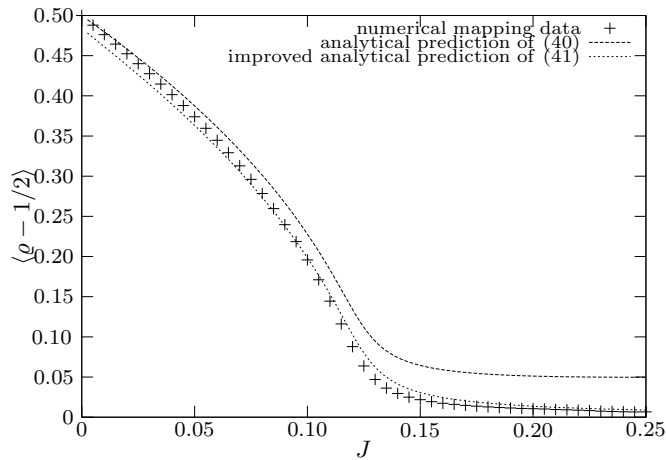}
\caption{Comparison of numerical data and analytical predictions for Lorentzian distribution of disorder with $\bar{\eta}=2.0$ and width=0.2.  Relatively large width means affect of $\eta < 0$ cutoff is significant.}
\label{f:lcomp2}
\end{center}
\end{figure}

From this stationary distribution one can calculate $\Delta$; averaging over $\varrho$'s from $-\infty$ to $+\infty$ gives
\begin{equation}
\Delta = \tfrac{1}{2} \Gamma x^{-1/2}.  \label{e:lorentzdel}
\end{equation}
As discussed above (see text following equation~\eqref{e:hcresJ}), this extension of the range of $\varrho$ is not expected to make much difference unless the current $J$ is large.  In fact for this Lorentzian case we can actually do the integral over $\varrho$ between 0 and 1 analytically, yielding the proper result
\begin{multline}
\Delta = \frac{x^{1/2}}{2\pi} \ln \left( \frac{(x^{1/2} - \Gamma )^2 +4x^2}{(x^{1/2} + \Gamma )^2 +4x^2} \right) \\ 
- \frac{\Gamma}{2\pi x^{1/2}} \, \arctan\left(\frac{4x^{3/2}}{\Gamma^2+4x^2-x}\right). \label{e:lorentzdelimp}
\end{multline}
In Fig.~\ref{f:lcomp2} we compare the results of these expressions with numerical data for a Lorentzian distribution of disorder and find
as expected that~\eqref{e:lorentzdelimp} gives a noticeably better fit than~\eqref{e:lorentzdel} for $J$ above the critical point.
The fit is still not exact due to the fact that in the numerics we have imposed the physical restriction that $\gamma$ must be positive but this cutoff is not incorporated in the analytics.  This problem is more pronounced than in the Gaussian case because of the relatively high weight in the tails of the Lorentzian distribution.

\section{Effect of disorder on fundamental and phase diagrams}
\label{s:fundpd}

We have seen that the shift $\Delta$ provides a useful characterization of the effect of disorder; we now turn our attention to what this shift means physically in terms of the density profiles, fundamental diagram (for periodic boundary conditions) and phase diagram (for open boundary conditions).   In Section~\ref{s:map} we presented both numerical and analytical approaches to calculate $\Delta$ for fixed $J$.  However in simulations $J$ is not held constant and we must consider how it changes when we add disorder.  We concentrate initially on the thermodynamic limit where $J$ is easier to predict and then in~\ref{ss:finite} consider the importance of finite size effects in small systems.

Recall that the treatment of Section~\ref{s:map} was based on using equation~\eqref{e:map} to map forwards in $\varrho$.  If instead we had used~\eqref{e:sigmap} to map backwards in $\sigma$ then by exactly the same argument we would have concluded that the steady state distribution for the disordered case has $\langle \sigma \rangle > 1/2$ i.e., $\langle \varrho \rangle < 1/2$ for all $J$.  This apparent paradox is exactly analogous to the situation in the pure low current phase where the stable fixed point value of $\varrho$ depends on which direction we map (see discussion in~\ref{ss:mapint}).  The resolution in both cases is that which fixed point is seen in the bulk depends on the boundary conditions.  We now consider these in more detail.

\subsection{Periodic boundary conditions}
\label{ss:fund}

Perhaps the most obvious change expected when we add disorder to a large system is a decrease in the maximum sustainable current.  For an infinitely large pure system with periodic boundary conditions the maximum current $J^0_{\text{max,p}}$ is just the critical current $J^0_C=p/4$.  Similarly for a disordered system, the maximum possible current is limited by stretches of ``weak'' bonds (i.e., low $p$) so in the thermodynamic limit we expect $J_{\text{max,p}}=J_C=p_\text{min}/4$ (where $p_\text{min}$ is the smallest value of $p$ permitted by our distribution of disorder).  Using the methods of Section~\ref{s:map}, we can obtain $\Delta$ corresponding to all possible currents up to $J_C$.  In contrast to the pure case we now have a non-zero value of $\Delta(J_C)$ which we shall denote for convenience by $\Delta_C$.  We now address how this is reflected in the fundamental diagram.

Let us first consider currents below the maximum.  Clearly, one possibility is for the distribution of $\varrho$ for all sites in the lattice to be given by the stationary distribution $w(\varrho)$ determined from the condition~\eqref{e:int}.  The average density is then obviously given by $1/2 + \Delta$.  However, the argument above illustrates that it's also possible for the distribution to be at the unstable fixed point of the forward density mapping giving $\langle \sigma - 1/2 \rangle = \Delta$ and hence an average density of $1/2 - \Delta$.   So, in the low current phase with periodic boundary conditions the possible disordered profiles are roughly the same as in the pure case (i.e., either at the upper fixed point or the lower fixed point) although of course with added noise.   The exact position of the fixed points (characterized by $\Delta$) deviates slightly from the pure case especially close to the critical point.

\begin{figure}
\begin{center}
\psfrag{r}[][]{$\varrho$}
\psfrag{1}[][]{1}
\psfrag{0}[][]{0}
\psfrag{A}[][]{$A$}
\psfrag{B}[][]{$B$}
\psfrag{C}[][]{$C$}
\psfrag{D}[][]{$D$}
\psfrag{E}[][]{$E$}
\psfrag{F}[][]{$F$}
\psfrag{l}[][]{$l$}
\psfrag{d1}[Cr][Cr]{{$1/2+\Delta_C$}}
\psfrag{d2}[Cr][Cr]{{$1/2-\Delta_C$}}
\includegraphics*[width=0.6\columnwidth]{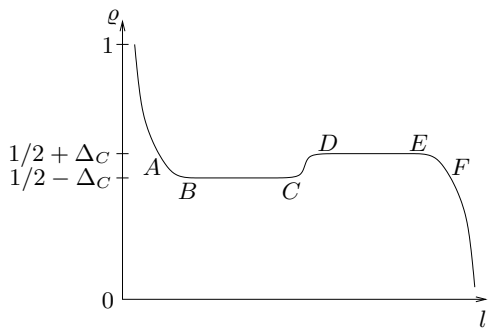}
\caption{Possible density profile in maximum current phase for system with disorder.  Periodic boundary conditions can be imposed for example between points $BC$, $AF$ or $DE$ leading to a range of possible densities from $1/2 - \Delta_C$ to $1/2 + \Delta_C$.}
\label{f:disprof}
\end{center}
\end{figure}

However, \emph{in the maximum current phase} it is possible for the distribution to start near the lower unstable fixed point and map forward to the upper stable fixed point via a noisy shock front (whose position may alter).  The periodic boundary conditions are maintained by stretches of decreasing $\varrho$ corresponding to weak bonds in the pure high current phase.  A crude way to consider this is to look at the profile as a superposition of a high current profile with a small shock~front type low current profile as shown schematically in Fig.~\ref{f:disprof}. 
This ``density segregation'' into sections (not necessarily of equal lengths) with density $1/2 - \Delta_C$ and $1/2 + \Delta_C)$ was previously explained for the binary disorder case by Tripathy and Barma~\cite{Tripathy98}.  Enforcement of the periodic boundary conditions (see again Fig.~\ref{f:disprof}) then leads to macroscopic average densities in this maximal current phase anywhere from $1/2 - \Delta_C$ to $1/2 + \Delta_C)$.

\begin{figure}
\begin{center}
\psfrag{J}[][]{$J$}
\psfrag{r}[][]{$\varrho$}
\psfrag{1}[][]{1}
\psfrag{0}[][]{0}
\psfrag{5}[Tc][Bc]{$\tfrac{1}{2}$}
\psfrag{p}[Cr][Cr]{$\tfrac{p_\text{min}}{4}$}
\psfrag{2D}[][]{$2\Delta_C$}
\includegraphics*[width=0.59\columnwidth]{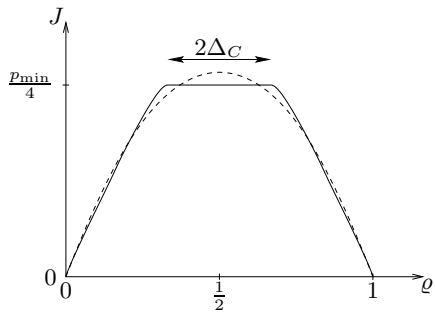}
\caption{Schematic fundamental diagram for DASEP in thermodynamic limit.}
\label{f:disfun}
\end{center}
\end{figure}

So the end result is a fundamental diagram which looks like the pure one for low currents but has a new flat regime of width 2$\Delta_C$ at the maximum current as shown in Fig.~\ref{f:disfun}.
This flattening effect was observed in~\cite{Tripathy98} and the width of the flat section calculated for the particularly simple case of binary disorder.  Our method enables us to treat more general distributions of disorder---from $\Delta(J)$ and the maximum current then we can construct the complete mean-field fundamental diagram.  

\begin{figure}
\begin{center}
\psfrag{J}[Bc][Tc]{$J$}
\psfrag{r}[Tc][Tc]{$\varrho$}
\psfrag{0}[Tr][Tr]{\footnotesize{0}}
\psfrag{0.2}[Tc][Tc]{\footnotesize{0.2}}
\psfrag{0.4}[Tc][Tc]{\footnotesize{0.4}}
\psfrag{0.6}[Tc][Tc]{\footnotesize{0.6}}
\psfrag{0.8}[Tc][Tc]{\footnotesize{0.8}}
\psfrag{1}[Tc][Tc]{\footnotesize{1.0}}
\psfrag{0.02}[Cr][Cr]{\footnotesize{0.02}}
\psfrag{0.04}[Cr][Cr]{\footnotesize{0.04}}
\psfrag{0.06}[Cr][Cr]{\footnotesize{0.06}}
\psfrag{0.08}[Cr][Cr]{\footnotesize{0.08}}
\psfrag{0.1}[Cr][Cr]{\footnotesize{0.10}}
\psfrag{0.12}[Cr][Cr]{\footnotesize{0.12}}
\psfrag{0.14}[Cr][Cr]{\footnotesize{0.14}}
\psfrag{"unDDfin5000fundt.dat"}[Cr][Cr]{\scriptsize{uniform disorder}}
\psfrag{"unpurefin5000fundt.dat"}[Cr][Cr]{\scriptsize{pure}}
\includegraphics*[width=1.0\columnwidth]{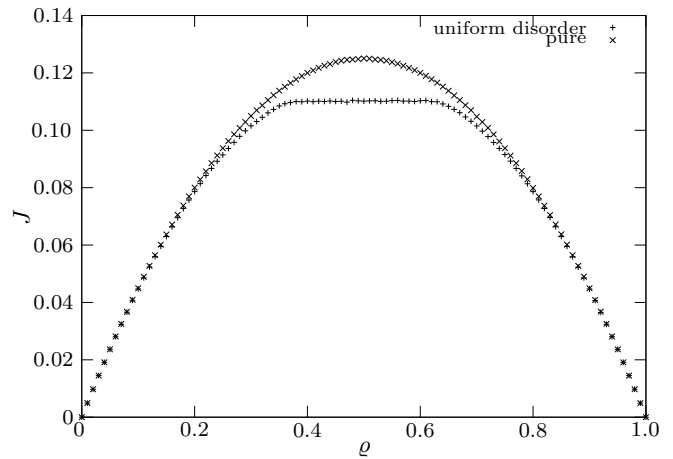}
\caption{Monte Carlo simulation data for DASEP fundamental diagram for system size 5000.  The uniform disorder case has $\bar{\eta}=2.0$ and $\sigma_\eta=0.5$; flattening with respect to the pure case is clearly seen.
}
\label{f:mcfund}
\end{center}
\end{figure}

Figure~\ref{f:mcfund} shows the fundamental diagram obtained from Monte Carlo simulations for a system of size 5000 with a particular realization drawn from a uniform distribution of disorder (results for the pure case corresponding to the mean of $1/p$ are also shown).  Comparison with the calculated $\Delta(J)$ for the same width of disorder (Fig.~\ref{f:distnum}) shows that the mean-field prediction is roughly correct, though the maximum current in the simulation is larger than the mean-field thermodynamic prediction of 0.1. 

\subsection{Open boundary conditions}
\label{ss:pd}

Popkov and Sch{\"u}tz~\cite{Popkov99} have shown how to predict the phase diagram for open boundary conditions from the fundamental diagram for periodic boundary conditions.  Their argument considers the motion of shock fronts thorough the bulk to motivate an extremal current principle:
\begin{alignat}{3}
J &= \underset{\varrho\in[\varrho^+,\varrho^-]}{\text{max}} J(\varrho) \quad &\text{for} \quad \varrho^- &> \varrho ^+ \\
J &= \underset{\varrho\in[\varrho^-,\varrho^+]}{\text{min}} J(\varrho) \quad &\text{for} \quad \varrho^- &< \varrho ^+
\end{alignat}
where $\varrho^-$ and $\varrho^+$ are reservoir densities (see~\ref{ss:model}).

\begin{figure}
\begin{center}
\psfrag{J}[][]{High $J$}
\psfrag{p1}[][]{Low $\varrho$}
\psfrag{p2}[][]{High $\varrho$}
\psfrag{B}[Cr][Cr]{$(1-\varrho^+)$}
\psfrag{A}[Tc][Tc]{$\varrho^-$}
\psfrag{5}[][]{$\tfrac{1}{2}$}
\psfrag{0}[][]{0}
\psfrag{1}[][]{1}
\psfrag{D1}[Bc][Bc]{$\Delta_C$}
\psfrag{D2}[Cl][Cl]{$\Delta_C$}
\includegraphics*[width=0.6\columnwidth]{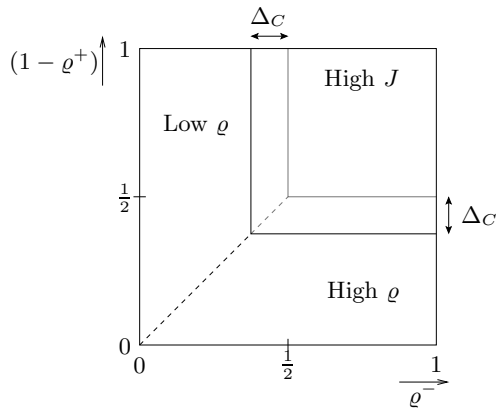}
\caption{Phase diagram for DASEP in the $\varrho^-$--$\varrho^+$ plane (black lines) with pure case for comparison (grey lines).}
\label{f:pd}
\end{center}
\end{figure}

\begin{figure}
\begin{center}
% Need to sort psfragging out for this.
\psfrag{J}[Bc][Tc]{$J$}
\psfrag{r1}[][]{$\varrho^-$}
\psfrag{r2}[Bc][Tc]{$\varrho^+$}
\psfrag{0}[Cr][Cr]{\scriptsize{0}}
\psfrag{0.1}[Cr][Cr]{\scriptsize{0.1}}
\psfrag{0.2}[Cr][Cr]{\scriptsize{0.2}}
\psfrag{0.3}[Cr][Cr]{\scriptsize{0.3}}
\psfrag{0.4}[Cr][Cr]{\scriptsize{0.4}}
\psfrag{0.5}[Cr][Cr]{\scriptsize{0.5}}
\psfrag{0.6}[Cr][Cr]{\scriptsize{0.6}}
\psfrag{0.7}[Cr][Cr]{\scriptsize{0.7}}
\psfrag{0.8}[Cr][Cr]{\scriptsize{0.8}}
\psfrag{0.9}[Cr][Cr]{\scriptsize{0.9}}
\psfrag{1}[Cr][Cr]{\scriptsize{1.0}}
\psfrag{0.02}[Cr][Cr]{\scriptsize{0.02}}
\psfrag{0.04}[Cr][Cr]{\scriptsize{0.04}}
\psfrag{0.06}[Cr][Cr]{\scriptsize{0.06}}
\psfrag{0.08}[Cr][Cr]{\scriptsize{0.08}}
\psfrag{0.1}[Cr][Cr]{\scriptsize{0.10}}
\psfrag{0.12}[Cr][Cr]{\scriptsize{0.12}}
\psfrag{0.14}[Cr][Cr]{\scriptsize{0.14}}
\includegraphics*[width=0.8\columnwidth]{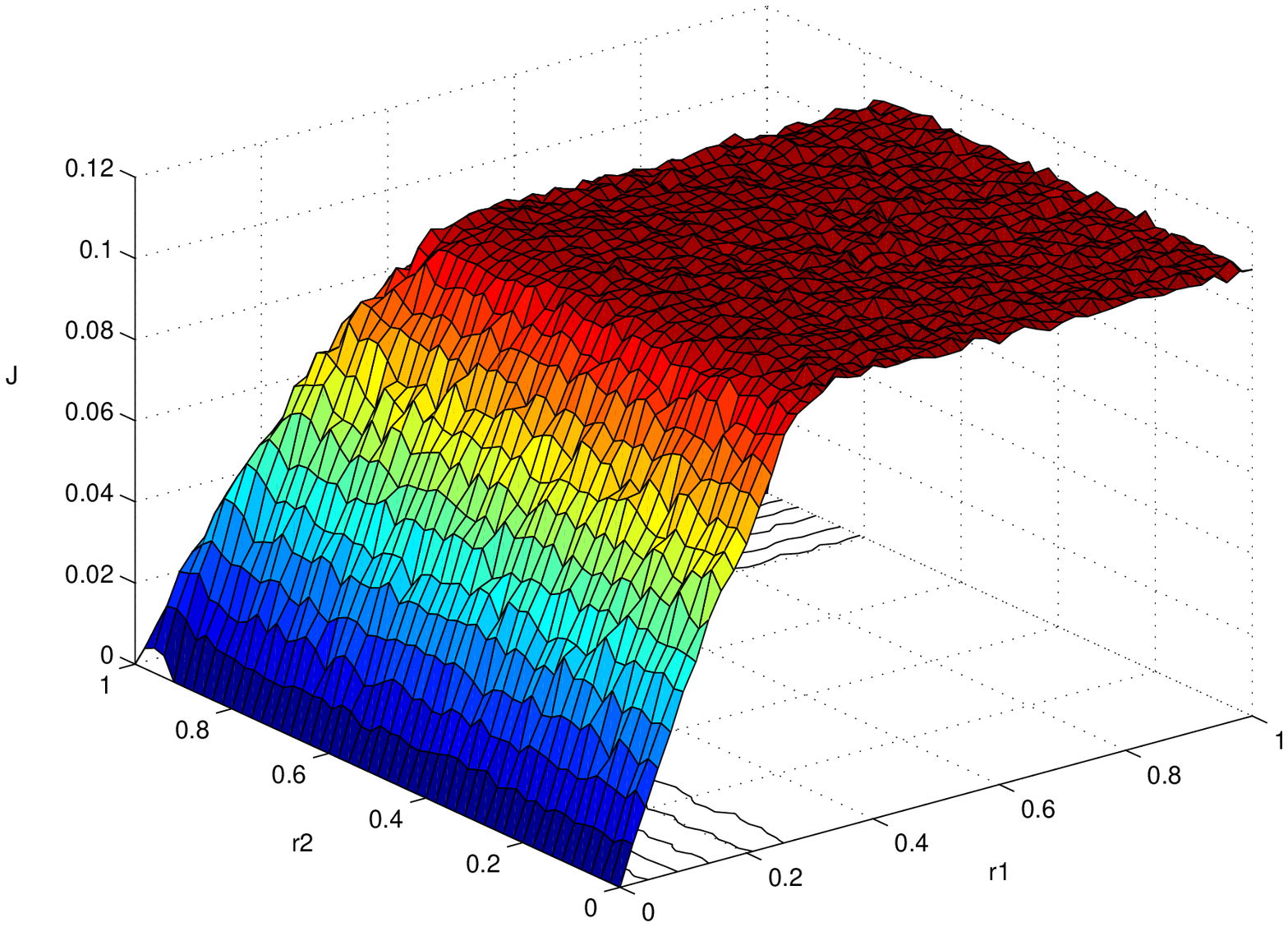}
\psfrag{0.1}[Cr][Cr]{\scriptsize{0.1}}
\includegraphics*[width=0.6\columnwidth, height=0.59\columnwidth]{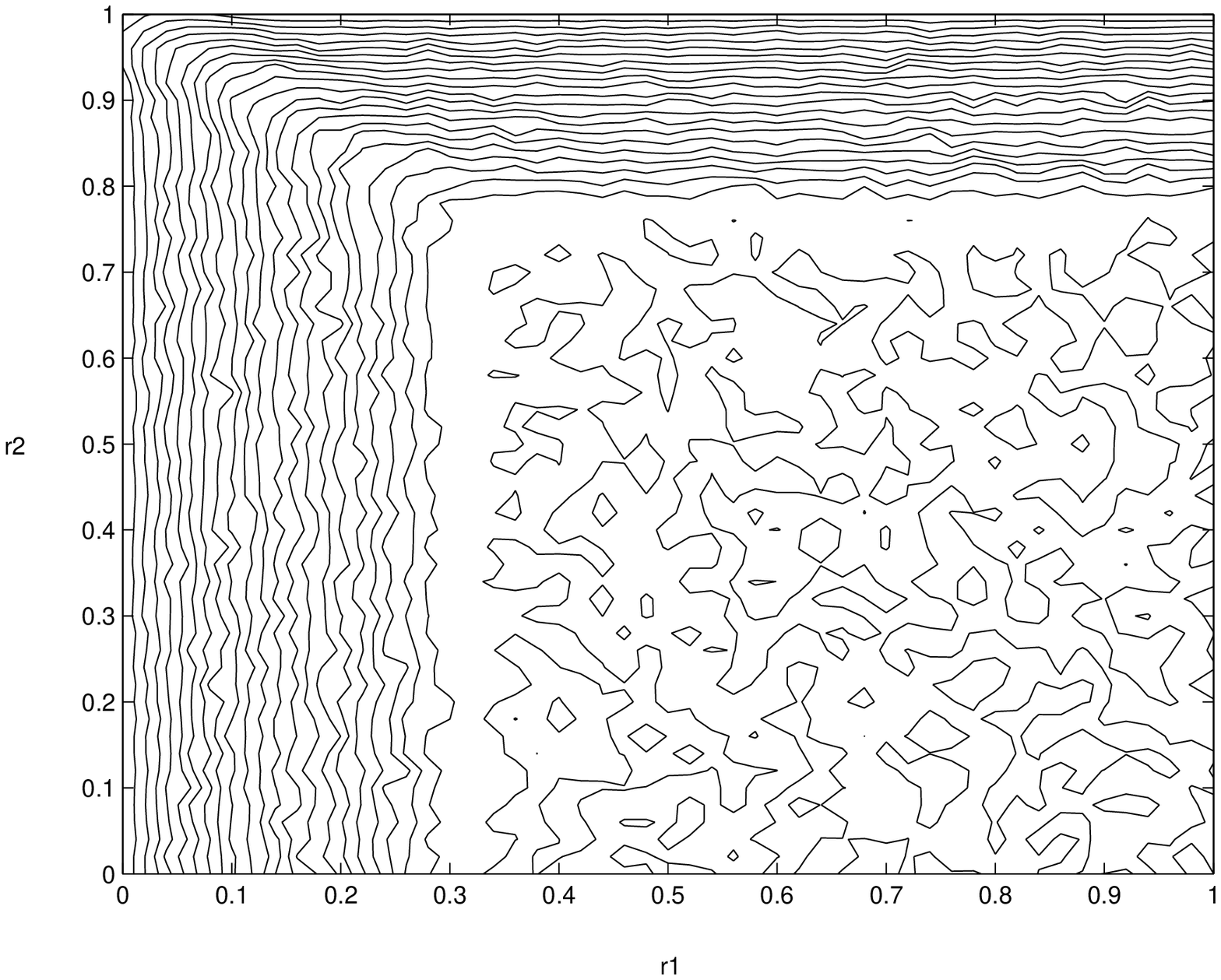}
\caption{(Color online).  Monte Carlo simulation results for DASEP phase diagram---surface and contour plots showing current as a function of $\varrho^-$, $\varrho^+$.  Uniform distribution with $\bar{\eta}=2.0$ and $\sigma_\eta=0.5$, system size 5000.  Note increase in size of flat maximum current area compared to the pure case. 
}
\label{f:pdmc}
\end{center}
\end{figure}

Applying this Popkov--Sch{\"u}tz argument to the fundamental diagram of~\ref{ss:fund} leads to a growth in the size of the high current phase (as compared to the pure case) resulting from the flat section on the fundamental diagram.  This growth is by an amount $\Delta_C$ in both the $\varrho^-$ and the $\varrho^+$ direction.  The resulting phase diagram in the $\varrho^-$--$\varrho^+$ plane and comparison with the pure case is shown schematically in Fig.~\ref{f:pd}.  
So our numerical and analytical calculations of $\Delta(J)$ in the previous section allow us to determine quantitatively the disordered phase diagram; compare the simulation data of Fig.~\ref{f:pdmc}.
\subsection{Finite Size Effects}
\label{ss:finite}

In this subsection we outline briefly the modifications to the above picture for finite size systems.   The discussion is inevitably fairly qualitative and it is worth noting that even in the pure case, a mean-field treatment does not correctly capture all finite size effects.  

In the pure case there are two main finite size effects.  Firstly in the open boundary case the system can sustain a current $J_\text{max,o}$ which is slightly larger than $J_C^0$.  One can obtain a mean-field prediction for $J_\text{max,o}$ by looking for the largest value of $J$ for which all $\varrho_l$ of~\eqref{e:mfprofhigh} are in the physically applicable regime between 0 and 1.  Secondly,  in a small system $J_C^0$ is increased slightly from $p/4$---this effect is due correlations between particle densities at adjacent sites and is therefore not reproduced by mean-field theory which predicts $J_C^0=p/4$ for all system sizes.   

In the disordered case $J_C$ is limited by stretches of weak bonds and the probability of finding within the system a long stretch of consecutive weak bonds increases with system size.  Hence in mean-field theory we expect $J_C$ to be greater than $p_\text{min}/4$ but less than the corresponding pure result.  It is relatively straightforward to calculate the expected mean-field $J_C$ for simple cases such as a binary distribution (see e.g.,~\cite{Tripathy98}).  However, just as in the pure case, we expect the true value of $J_C$ for small systems to be larger than this mean-field estimate.  In addition, a novel feature of the disorder is that it allows density profiles in the periodic boundary case such as that shown in Fig.~\ref{f:disprof} where the current flow is larger than $J_C$, i.e., for finite disordered systems we can have $J_\text{max,p} > J_C$.  It is even possible to conceive of situations in which $J_\text{max,p}$ is increased above $J_C^0$ by adding disorder, meaning that $\Delta$ above the pure critical point (as calculated in~\ref{ss:above}) can be a physically relevant quantity.  
Examination of the possible high current profiles reveals that for small systems the current varies with macroscopic density (with the maximum $J_\text{max,p}$ at $\varrho=0.5$) so a flat section is not seen in the fundamental diagram.  Similarly in the disordered finite size open boundary case, $J_{\text{max,o}}$ is increased beyond $J_C$ and there are alterations in the phase diagram corresponding to the altered fundamental diagram.

So finite size effects have a significant complicating influence on both the fundamental diagram and the phase diagram.  At present we are not able to quantify these entirely even within mean-field but progress can be made by combining numerical work (e.g., self-consistently looking for the maximum current the density mapping can sustain) with the analysis of previous sections.

\section{Results from disordered Cole--Hopf transformation}
\label{s:colehopf}

Recall from Subsection~\ref{ss:chintro} that we are able to treat the continuum limit of the DASEP via a disordered generalization of the Cole--Hopf transformation~\eqref{e:invCH} to obtain the linear equation~\eqref{e:disdiff}.  Here we develop this approach further and demonstrate connections to the results from the steady state mapping.

\subsection{Scaling and localization in the steady state}

Let us concentrate initially on the steady state solution for $\varrho$ in order to make comparisons with the discrete mean-field mapping approach.  As discussed in~\ref{ss:chintro}, a steady state solution for $\varrho$ corresponds to a separable solution for $u$ i.e., $u(x)= T(t)X(x)$.  The $x$-dependent factor $X$ must then satisfy
\begin{equation}
-[\omega - D(x)]X = D(x) \frac{d^2 X}{dx^2}
\end{equation}
with $\omega=2J$.

Our approach is to rediscretize this,
\begin{equation}
(3D_n-\omega)X_n=D_n(X_{n+1}+X_{n-1}) \label{e:dislin},
\end{equation} 
and then employ numerical scaling based on a method developed by Pimentel and Stinchcombe~\cite{Pimentel88} to treat the equation of motion of a 1D Mattis-transformed Edwards-Anderson Heisenberg spin glass.  We write~\eqref{e:dislin} as
\begin{equation}
(E_n- \zeta_n \omega) X_n = V_{n,n-1} X_{n-1} + V_{n,n+1} X_{n+1} \label{e:mattis}
\end{equation}
with $\zeta_n = D/D_n$, $E_n=3D$, $V_{n,n+1}=D$ where $D$ is the characteristic strength of the disorder variable.  Equation~\eqref{e:mattis} is of just the form considered in~\cite{Pimentel88} and can be exactly scaled by $b=2$ decimation to give
\begin{equation}
(E'_n- \zeta_n \omega) X_n = V'_{n,n-2} X_{n-2} + V'_{n,n+2} X_{n+2} \label{e:mattis2}
\end{equation}
with
\begin{align}
V'_{n,n+2} &= \frac{V_{n,n+1} V_{n+1,n+2}}{E_{n+1} - \zeta_n \omega} \\
E'_n &= E_n - \frac{V^2_{n,n-1}}{E_{n-1} - \zeta_{n-1}\omega} - \frac{V^2_{n,n+1}}{E_{n+1} - \zeta_{n+1}\omega}.
\end{align}
So the $E_n$ and $V_{n,n\pm1}$, which are n-independent at the outset, pick up correlated randomness under scaling.

It is easy to iterate these equations numerically and check how the $V$'s evolve.  For the pure case (i.e., all $\zeta_n=1$) we find an allowed energy band for $D<\omega<5D$.   Within this band the ``site potential'' $E$ and the ``coupling'' $V$ evolve chaotically, corresponding to extended states.  Outside the band there are no allowed states and $V$ decreases rapidly and monotonically to zero while $E$ tends to a constant value.  Just as in ~\cite{Pimentel88} this can be explained by writing $V$ and $E$ explicitly in terms of a single parameter $\theta$ which is related to the wave vector of excitations within the band.  As expected the lower edge of the allowed band $\omega=D$, corresponds to the pure critical current $J_C^0=p/4$.  The upper band edge has no physical significance in our problem since this switch between continuum and discrete representations is valid only for long wavelengths corresponding to being close to $\omega=p/4$. 

\begin{figure}
\begin{center}
\psfrag{w}[][]{$\omega$}
\psfrag{l}[Bc][Tc]{$\xi$}
\psfrag{0}[Tr][Tr]{\footnotesize{0}}
\psfrag{0.2}[Tc][Tc]{\footnotesize{0.2}}
\psfrag{0.4}[Tc][Tc]{\footnotesize{0.4}}
\psfrag{0.6}[Tc][Tc]{\footnotesize{0.6}}
\psfrag{0.8}[Tc][Tc]{\footnotesize{0.8}}
\psfrag{1}[Tc][Tc]{\footnotesize{1.0}}
\psfrag{1.2}[Tc][Tc]{\footnotesize{1.2}}
\psfrag{1.4}[Tc][Tc]{\footnotesize{1.4}}
\psfrag{1.6}[Tc][Tc]{\footnotesize{1.6}}
\psfrag{50}[Cr][Cr]{\footnotesize{50}}
\psfrag{100}[Cr][Cr]{\footnotesize{100}}
\psfrag{150}[Cr][Cr]{\footnotesize{150}}
\psfrag{200}[Cr][Cr]{\footnotesize{200}}
\psfrag{"pure.dat"}[Cr][Cr]{\scriptsize{pure, $\sigma_\eta=0$}}
\psfrag{"DgaussNEW.dat"}[Cr][Cr]{\scriptsize{disorder with $\sigma_\eta=0.2$}}
\includegraphics*[width=1.0\columnwidth]{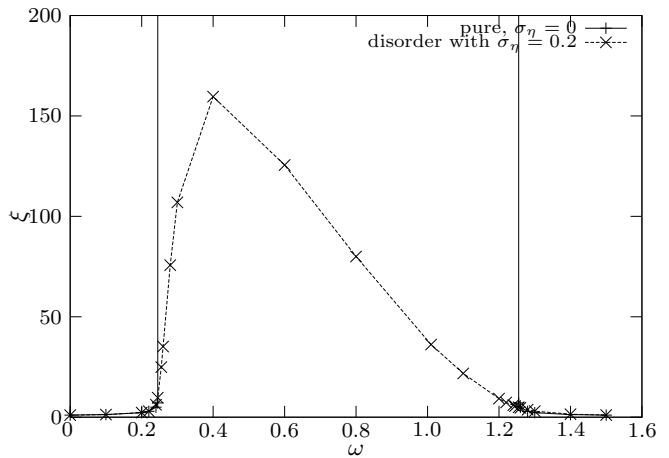}
\caption{Localization transition in Cole--Hopf mapped DASEP.  The data points represent the localization length $\xi$ calculated by our numerical scaling method; lines are provided as an aid to the eye.  In the pure case (with $\bar{\eta}=2.0$) the localization length is effectively infinite in the ``allowed band'' $0.25<\omega<1.25$; disorder is seen to induce localization.}  
\label{f:loclength}
\end{center}
\end{figure}

Adding weak randomness, we find that for all values of $\omega$, $V$ evolves either chaotically or cyclically to zero inside a well defined exponentially decreasing envelope i.e., $V(l) \sim f(l) e ^{-l/\xi}$ where $l$ is the distance between sites and $\xi$ is a localization length.  In other words, any amount of disorder induces localization for all values of frequency.  This is analogous to the fact that all states are localized in one-dimensional disordered quantum problems.  We developed a computer algorithm to calculate  $\xi$ for a given distribution of disorder (averaging over many realizations) as a function of $\omega$.  Typical results for both pure and disordered cases are shown in Fig.~\ref{f:loclength}.  
Note that, in contrast to the spin-glass case studied in~\cite{Pimentel88}, the localization length is not infinite at the critical point.  This is essentially due to the $-D(x) u$ term in \eqref{e:disdiff} and means that we cannot easily follow~\cite{Pimentel88} in defining a dynamic exponent via a relationship like $\omega \propto (1/\xi)^z$.

To determine the signature of localization in the original DASEP problem we consider a localized form of $u$,
\begin{equation}
u \sim e^{\pm x/\xi}.
\end{equation}
Using~\eqref{e:invCH} to invert the Cole--Hopf transformation gives
\begin{equation}
\varrho = \frac{1}{2} \pm \frac{1}{2\xi}. \label{e:delxi}
\end{equation}
i.e., a shift in the profile.  So calculating the average localization length is just like calculating the average shift from $\varrho=1/2$ in the original problem.  In other words the average localization length is just half the inverse of the quantity $\Delta$ defined in Section~\ref{s:map}. Of course, in general $u$ will be some boundary-dependent combination of $e^{+x/\xi}$ and $e^{-x/\xi}$, corresponding to the fact that, in the high current phase, the observed shift in density can be anywhere between $+\Delta$ and $-\Delta$ i.e., the flat section on the fundamental diagram.  This relief of localization in the inverse mapping is possibly connected with the work of Kopidakis and Aubry~\cite{Kopidakis00b} on the relief of localization by non-linearity in low-dimensional deterministic systems.

\begin{figure}
\begin{center}
\psfrag{w}[][]{$\omega$}
\psfrag{l}[][]{$\xi$}
\psfrag{0}[Tr][Tr]{\footnotesize{0}}
\psfrag{0.2}[Tc][Tc]{\footnotesize{0.2}}
\psfrag{0.4}[Tc][Tc]{\footnotesize{0.4}}
\psfrag{0.6}[Tc][Tc]{\footnotesize{0.6}}
\psfrag{0.8}[Tc][Tc]{\footnotesize{0.8}}
\psfrag{1}[Tc][Tc]{\footnotesize{1.0}}
\psfrag{1.2}[Tc][Tc]{\footnotesize{1.2}}
\psfrag{1.4}[Tc][Tc]{\footnotesize{1.4}}
\psfrag{1.6}[Tc][Tc]{\footnotesize{1.6}}
\psfrag{50}[Cr][Cr]{\footnotesize{50}}
\psfrag{100}[Cr][Cr]{\footnotesize{100}}
\psfrag{150}[Cr][Cr]{\footnotesize{150}}
\psfrag{200}[Cr][Cr]{\footnotesize{200}}
\psfrag{250}[Cr][Cr]{\footnotesize{250}}
\psfrag{"DgaussNEW.dat"}[Cr][Cr]{\scriptsize{numerical data for $\xi$}}
\psfrag{"mapegbigrangebwloc2.dat"}[Cr][Cr]{\scriptsize{numerical data for $(2\Delta)^{-1}$ }}
\psfrag{g(x)}[Cr][Cr]{\scriptsize{analytical expression for $\xi$}}
\psfrag{f(x)/2}[Cr][Cr]{\scriptsize{analytical expression for $(2\Delta)^{-1}$}}
\includegraphics*[width=1.0\columnwidth]{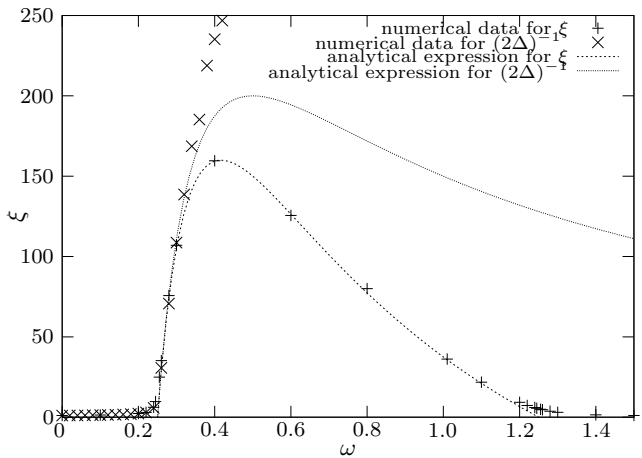}
\caption{Comparison of localization length $\xi$ obtained by numerical scaling with $(2\Delta)^{-1}$ from the mapping approach.  Disorder with $\bar{\eta}=2.0$, $\sigma_\eta=0.2$.  Analytical predictions for $\xi$ and $(2\Delta)^{-1}$ in the high current regime are also shown.}  
\label{f:prepostcomp2}
\end{center}
\end{figure}

In Fig.~\ref{f:prepostcomp2} 
we explicitly compare the localization length obtained by this scaling method and $(2\Delta)^{-1}$ from the discrete mapping method.  We find an excellent agreement for $J$ near to the critical point where the discrete--continuum--discrete approximations are valid, but the comparison breaks down in the high $J$ region which in any case is unphysical.  Furthermore, by defining $Y_{n+1}=X_{n+1}/X_n$ we can cast~\eqref{e:dislin} into exactly the form studied by Hirota~\cite{Hirota73}
\begin{equation}
Y_{n+1} = \alpha_n - 1 / Y_n \label{e:hmap}
\end{equation}
with 
\begin{equation}
\alpha_n=(3-\zeta_n \omega).
\end{equation}
We can then obtain the stationary probability distribution $w(Y)$ in analogy with the calculation of Section~\ref{s:map} and, following~\cite{Hirota73}, define the localization length $\xi$ by
\begin{equation}
\frac{1}{\xi} = \frac{1}{2} \int^{\infty}_{-\infty} w(Y) \ln Y^2 \, dY
\end{equation}
leading eventually to
\begin{equation}
\xi = \frac{[4-(3-2\bar{\eta}\omega)^2]}{2\omega^2 \sigma_\eta^2}.
\end{equation}

The resulting prediction for the localization length of the high current phase agrees closely with that determined by our numerical scaling method (see Fig.~\ref{f:prepostcomp2}) except near the critical point where the Hirota method breaks down (just as in the pre Cole--Hopf case).  One subtlety involves the meaning of ``averaging over realizations'' in finding the average localization length.  In the original numerical scaling method of~\cite{Pimentel88} the averaging is over $V \sim e^{-l/\xi_i}$ (where $i$ labels the specific realization of disorder); in contrast the Hirota method takes the average of $1/\xi_i$.   It is clear from~\eqref{e:delxi} that averaging over $1/\xi_i$ provides the definition of $\xi$ most directly comparable with $\Delta$ so this is the procedure adopted in the computer programs used to generate the data in Figures~\ref{f:loclength} and~\ref{f:prepostcomp2}.  Averaging over $V$ instead gave less good agreement with the analytical expression especially in the center of the band where the localization length is less clearly defined.

Finally we note that since the variable $Y$ can physically take all values between $-\infty$ and $+\infty$ we expect better agreement in the numerical and analytical results for $\xi$ than the numerical and analytical results for $\Delta$ where the analytics for general distributions of disorder was unable to take account of the physical restriction $0 \le \varrho \le 1$ resulting in discrepancies for high values of $J$ (see discussion of~\ref{ss:above}).  In addition, it is easier to get high quality data from the linear scaling computer algorithm (where we can measure accurately localization lengths up to $\sim$10,000) than from the non-linear mapping.

\subsection{Effect of disorder on dynamics}

One advantage of the Cole--Hopf formalism is that it provides an easy route to discuss the influence of disorder on the dynamics of the system.  Here we mention briefly some inferred effects.  A general solution for $u$ is made up of a superposition of separable solutions; in the pure case this leads to a general solution for $\varrho$ of the form, 
\begin{equation}
\varrho(x)-\frac{1}{2}= \frac{1}{2} \frac{\Sigma_k A_k ik e^{ikx-\omega_k t}}{\Sigma_k A_k e^{ikx-\omega_k t}} \label{e:rhogen}
\end{equation}
where $k$ can be positive or negative and is real in the high current phase and imaginary in the low current phase.  The coefficients ${A_k}$ must be chosen so that $\varrho$ is real.  From~\eqref{e:disdiff} the pure ``dispersion relation'' is $\omega_k = D(1 + k^2)$ but the $k$-independent term will cancel out in the numerator and denominator of~\eqref{e:rhogen}.  The solutions hence have wavelike form in the high current phase and multiple soliton form in the low current phase.  As time increases, transients die away leaving the steady state corresponding to the smallest value of $\omega$.

\begin{figure}
\begin{center}
\psfrag{r}[Bc][Tc]{$\varrho$}
\psfrag{t}[][]{$t$}
\psfrag{0}[Tc][Tc]{\footnotesize{0}}
\psfrag{5000}[Tc][Tc]{\footnotesize{5000}}
\psfrag{10000}[Tc][Tc]{\footnotesize{10000}}
\psfrag{15000}[Tc][Tc]{\footnotesize{15000}}
\psfrag{20000}[Tc][Tc]{\footnotesize{20000}}
\psfrag{25000}[Tc][Tc]{\footnotesize{25000}}
\psfrag{30000}[Tc][Tc]{\footnotesize{30000}}
\psfrag{35000}[Tc][Tc]{\footnotesize{35000}}
\psfrag{40000}[Tc][Tc]{\footnotesize{40000}}
\psfrag{0.4}[Cr][Cr]{\footnotesize{0.4}}
\psfrag{0.5}[Cr][Cr]{\footnotesize{0.5}}
\psfrag{0.6}[Cr][Cr]{\footnotesize{0.6}}
\psfrag{0.7}[Cr][Cr]{\footnotesize{0.7}}
\psfrag{0.8}[Cr][Cr]{\footnotesize{0.8}}
\psfrag{0.9}[Cr][Cr]{\footnotesize{0.9}}
\psfrag{1}[Cr][Cr]{\footnotesize{1.0}}
\psfrag{"pure500sprhotot.dat"}[Cr][Cr]{\scriptsize{pure}}
\psfrag{"dis500sprhotot.dat"}[Cr][Cr]{\scriptsize{disordered}}
\includegraphics*[width=1.0\columnwidth]{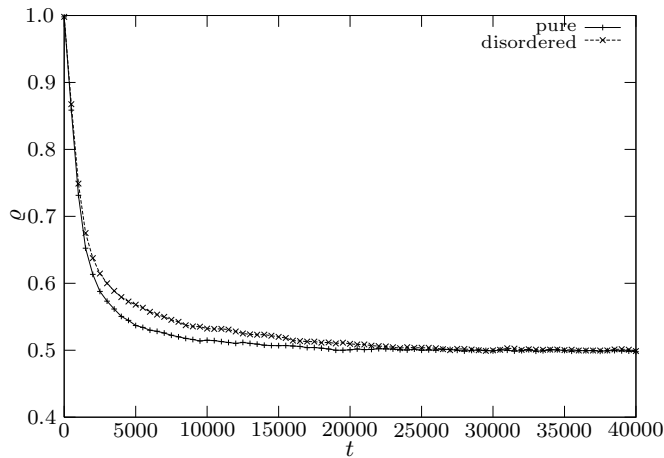}
\caption{Total density versus time in a system of 500 lattice sites fully occupied at $t=0$, reservoir densities $\varrho^-=1$, $\varrho^+=0$.  Disordered case is average over 100 realizations drawn from  a Gaussian distribution with $\bar{\eta}=2.0$ and $\sigma_\eta=0.5$.  A disorder-induced slow-down in the decay to the steady state ($\varrho=0.5$) is clearly observed; presumably any oscillations in the approach to the steady state are averaged out.}
\label{f:dyn}
\end{center}
\end{figure}

Disorder induced localization in the high current phase, is crudely like adding a small imaginary part $i\kappa$ ($\kappa \sim 1/\xi$) to the real $k$ (in~\eqref{e:rhogen} this gives steady state solutions for $\varrho$ like that in Fig.~\ref{f:disprof}).  The imaginary part of the resulting complex dispersion relation $\omega \sim D(1 + k^2-\kappa^2+2 i k \kappa)$ would be expected to lead to oscillations while the small decrease in the real part slows down the dynamics.  Indeed, it is intuitively obvious that adding disorder should slow down the approach to the steady state, since in a 1D system the overall rate of hopping of the particles (and hence the speed at which the steady state is reached) will be limited by the bond with the smallest $p_l$.  This slow-down is confirmed by the Monte Carlo simulations of Fig.~\ref{f:dyn}.  
 Within this framework a more quantitative analysis should be possible but would be complicated by boundary conditions and finite size effects (compare Section~\ref{s:fundpd}).

\section{Discussion and outlook}
\label{s:conc}

In this paper we have shown within a mean-field framework that one effect of quenched bond disorder on the ASEP is a flattening of the top of the steady state current--density relation and a corresponding increase in the high current region of the phase diagram for open boundary conditions.  We have presented various numerical and analytical approaches (including a mapping to a localization transition in an equivalent problem) to quantify these changes and shown that our results compare reasonably well with Monte Carlo simulations.

While we believe that this mean-field discussion reproduces qualitatively the effects of adding disorder, an exact treatment would be expected to provide better quantitative agreement with simulations together with further physical understanding.  Some progress has already been made (see the review by Stinchcombe~\cite{Stinchcombe02}).  For example a form of the Harris Criterion~\cite{Harris74} can be applied to the ASEP which suggests that disorder should be relevant in the sense of introducing new critical behavior.  And this new critical behavior can be elucidated by adding disorder to renormalization schemes developed for the pure case (see e.g.,~\cite{Georgiev02,Stinchcombe02b}).   In carrying out such scaling of distributions one re-encounters many of the concepts highlighted above such as Griffith's phases and the importance of tails of the distribution.

Many of the ideas discussed (e.g., Griffiths's phases, localization, importance of boundary conditions) might also be expected to apply to other nonequilibrium situations and it would be interesting to see if we can apply generalizations of the methodology of Sections~\ref{s:approach}--\ref{s:colehopf} to other problems.  In particular we have studied a simple two-lane traffic model~\cite{MeUnpub} in which the fundamental diagram can have a double-peak structure (the ASEP with next-nearest neighbor interactions as studied by Popkov and Sch{\"u}tz~\cite{Popkov99} also has such a double maxima).  We would expect that disorder flattens the tops of these maxima leading to corresponding changes in the phase diagram but more interesting effects are also possible such as a relative change in the height of the two peaks.  Studying quasi-1D models such as this two-lane system might also provide a bridge to understanding the effects of disorder on higher dimensional systems where one expects to find a wider range of possible disorder-induced effects.

In conclusion, we hope that this mean-field treatment of the disordered asymmetric simple exclusion process provides a flavor of the general phenomena present in nonequilibrium models with quenched substitutional disorder.  There is much scope for further work on this and related models.

\begin{acknowledgments}
We thank Kimmo Kaski for the Monte Carlo simulation data in Fig.~\ref{f:mapsim}.  This work was financially supported by EPSRC under Oxford Condensed Matter Theory Programme Grant GR/M04426/01 and Studentship 309261.
\end{acknowledgments}

\bibliographystyle{apsrev}
\bibliography{/home/wytham/harris/TeX/References/allref}

\end{document}